# Could face-centered cubic titanium in cold-rolled commercially-pure titanium only be a Ti-hydride?

Yanhong Chang [a], Siyuan Zhang [a*], Christian H. Liebscher [a], David Dye [b], Dirk Ponge [a], Christina Scheu [a], Gerhard Dehm [a], Dierk Raabe [a], Baptiste Gault [a,b*], Wenjun Lu [a*]

[a] *Max-Planck-Institut für Eisenforschung GmbH, Max-Planck-Straße 1, 40237, Düsseldorf, Germany*

[b] *Department of Materials, Royal School of Mines, Imperial College, Prince Consort Road, London SW7 2BP, United Kingdom*

Correspondence to: w.lu@mpie.de; siyuan.zhang@mpie.de; b.gault@mpie.de

**Abstract**

A face-centered cubic (FCC) phase in electro-polished specimens for transmission electron microscopy of commercially pure titanium has sometimes been reported. Here, a combination of atom-probe tomography, scanning transmission electron microscopy and low-loss electron energy loss spectroscopy is employed to study both the crystal structural and chemical composition of this FCC phase. Our results prove that the FCC phase is actually a $TiH_x$ ($x \geqslant 1$) hydride, and not a new allotrope of Ti, in agreement with previous reports. The formation of the hydride is discussed.

Ti and Ti-alloys are known to exhibit multiple crystallographic structures: a low-temperature α-hexagonal close packed phase (hcp), a high-temperature β-body-centered cubic phase (bcc), a hcp martensite α' or orthorhombic martensite α'', and a high-pressure metastable trigonal or hexagonal phase ω. A face-centered cubic (FCC) crystal structure of Ti has been sporadically reported from simulations [1,2] and experiments [3–12]. First-principles calculation [1] indicated that the total energy of Ti along the Bain transformation shows a minimum at a FCC structure. Aguayo et al. [2] suggested that HCP metals, e.g., Ti, Zr and Hf, have a locally stable FCC structure in a certain pressure regime from first-principles, full-potential calculations. Earlier experimental studies claimed that FCC-Ti phase exists in Ti thin films [3–6], high-energy ball milled Ti powder [7], Ti/Ni and Ti/Al multilayers [8–10] and bulk commercially pure Ti and Ti-alloys [12–23]. However, doubts were cast on some of these reports that the FCC-phase is associated to an artifact caused during preparation of transmission electron microscopy (TEM) specimens [24–31].

The FCC phase in bulk pure Ti and Ti-alloys appeared lath- or needle-shaped, its lattice parameter falls within the range of 4.1 – 4.4 Å [13,16–19,21–23], and the orientation relationship (OR) with the adjacent HCP matrix is reported to be either OR1: $<0001>_{HCP} // <001>_{FCC}$ ; $\{1\bar{1}00\}_{HCP} // \{1\bar{1}0\}_{FCC}$ [13,18,20,22,23]; or OR2: $<\bar{1}2\bar{1}0>_{HCP} // <1\bar{1}0>_{FCC}$; $\{0002\}_{HCP} // \{111\}_{FCC}$ [19,23]. Coincidentally, both the morphology and the crystallography of the FCC-Ti phase, determined by TEM/high-resolution TEM (HRTEM), are similar to those of the hydrides in Ti and its alloys [30,32–35]. Hydrides readily form in Ti, Zr, Hf and their respective alloys, because of a high affinity to hydrogen combined with low hydrogen solubility in the HCP α phase at room temperature. Hydrides were reported to form in Ti, Zr and some of their alloys during traditional specimen preparation processes, e.g., electropolishing [29,30], broad ion beam milling [24,36] or conventional focused ion beam (FIB) at room temperature [37–39]. Despite clear evidence from past literature on the influence of hydrides generated through TEM sample preparation in Ti-based alloys [27–31], the possibility that the FCC-phase could be a hydride has not been sufficiently appreciated in recent years.

The characterization of the H distribution in engineering alloys at the nano-scale is challenging. TEM has been used to image H-containing atomic columns in ordered phases [40], yet the direct imaging of individual light atoms is usually out of reach. However, hydrogen can be detected via plasmon excitation in electron energy loss spectroscopy (EELS), where hydrogen in metal hydrides usually introduces an upward shift in the plasmon peak energy compared to the parent metals [41–46]. An alternative is to use atom probe tomography (APT). APT is a mass spectroscopy technique with sub-nanometer spatial resolution, which enables the characterization and visualization of the 3D distribution of elements. With great care, H can be specifically analyzed in metallic materials, e.g., steels [47,48], Zr alloys [49–51] and Ti alloys [39,52].

Here, needle-shaped FCC-precipitates were observed in specimens for TEM of CP-Ti before and after cold-rolling, prepared by electropolishing at -30°C. We use (S)TEM-EELS to analyze the crystal structure and chemical composition in these precipitates. Specimens for APT were prepared by site-specific lift-out of FCC laths from within TEM specimens. The final stage of the preparation was performed at cryogenic temperature. This avoids significant hydrogen ingress [46,53], which was attributed to a decrease in the inwards diffusivity of H formed at the bare metallic surface during the preparation. The correlative STEM-APT provides the first direct evidence that hydrogen is present in these FCC phase, which we prove are Ti-hydride and not a new allotrope of Ti.

We used CP-Ti (Grade 2) in its as-received state and after cold-rolling at room temperature to a total thickness reduction of 30%. The hydrogen content in the bulk measured from thermal conductivity is approx. 30 wppm. TEM specimens, with a diameter of 3 mm, were mechanically ground to approx. 100μm and then thinned by twin jet-electropolishing in a solution of 6% perchloric acid + 59% methanol + 35% Butoxyethanol at −30 °C at an applied voltage of 15V. We performed TEM and selected area electron diffraction (SAED) by using a Titan Themis 60-300 (Thermo Fisher Scientific) microscope operated at an accelerating voltage of 300 kV. An aberration-corrected probe with semi-convergence angle of 24 mrad was used for STEM imaging and spectroscopy. For high-angle annular dark field (HAADF) and low angle annular dark field (LAADF) imaging, we used inner and outer semi-collection angles ranging from 73 to 200 mrad and 14 to 63 mrad, respectively. EELS spectra were recorded using a Quantum ERS (Gatan) energy filter in the image-coupled mode with an entrance aperture collecting electrons scattered up to 35 mrad. Multivariate analysis [54] was performed on the hyperspectral datasets to separate spectral contribution from the two phases.

HAADF-STEM imaging in Fig. 1a reveals numerous ~25 nm wide needle-shaped particles. Fig. 1b is a close-up on the region delineated by the red dashed square in Fig. 1a. The needle-shaped phase appears darker, compared to the surrounding matrix, indicating that this phase exhibits a lower average atomic number. LAADF-STEM imaging (Fig. 1c), strong strain contrast stemming from the high dislocation density after cold rolling is observed in the HCP matrix. Figs. 1d-f show atomic resolution HAADF-STEM images of the FCC, HCP phase, and their interface, respectively. Based on the fast-Fourier transform (FFT) analysis (insets in Figs. 1d-f), the orientation relationship between the FCC phase and the HCP matrix is OR1: $<0001>_{HCP} // <001>_{FCC}$ ; $\{01\bar{1}0\}_{HCP} // \{220\}_{FCC}$ . Meanwhile, four superlattice reflections

(marked by yellow arrows) in inset Fig. 1d indicate ordering of atoms in the FCC phase. This may be caused by an ordered arrangement of H atoms in the FCC lattice [55], which should result an ordered FCT hydride phase. However our measurements of the lattice parameters do not perfectly coincide with reported values for the FCC- or FCT-hydrides, which can be caused by either the fact that the observed hydrides are under strain in the Ti-lattice or may be an intricate mixture of different hydride phases. Apart from these needles, lenticular shape particles, approx. 80 nm in width, also with an FCC-structure, are observed in Fig. 2a-c. Based on the inset SAED in Fig. 2c, the orientation relationship between lenticular shaped FCC phase and matrix is confirmed to be OR2: $<\bar{1}2\bar{1}0>_{HCP} // <1\bar{1}0>_{FCC}$ ; $\{0002\}_{HCP} // \{111\}_{FCC}$. Besides, the lattice constant of both types of FCC phases is calculated to be approx. 0.42 nm, which is consistent with previously reported FCC phase in cold-rolled Ti metal and alloys [55]. Additional thin foils were prepared by in-situ lift-out on a FEI Helios PFIB equipped with a cold stage [56] and observed by transmission-Kukuchi diffraction on the same instrument with an EDAX HIKARI1 detector. None of these precipitates were observed by TKD following preparation of specimens by cryo-PFIB as shown in Fig. S1. or by TEM as reported in ref. [53]

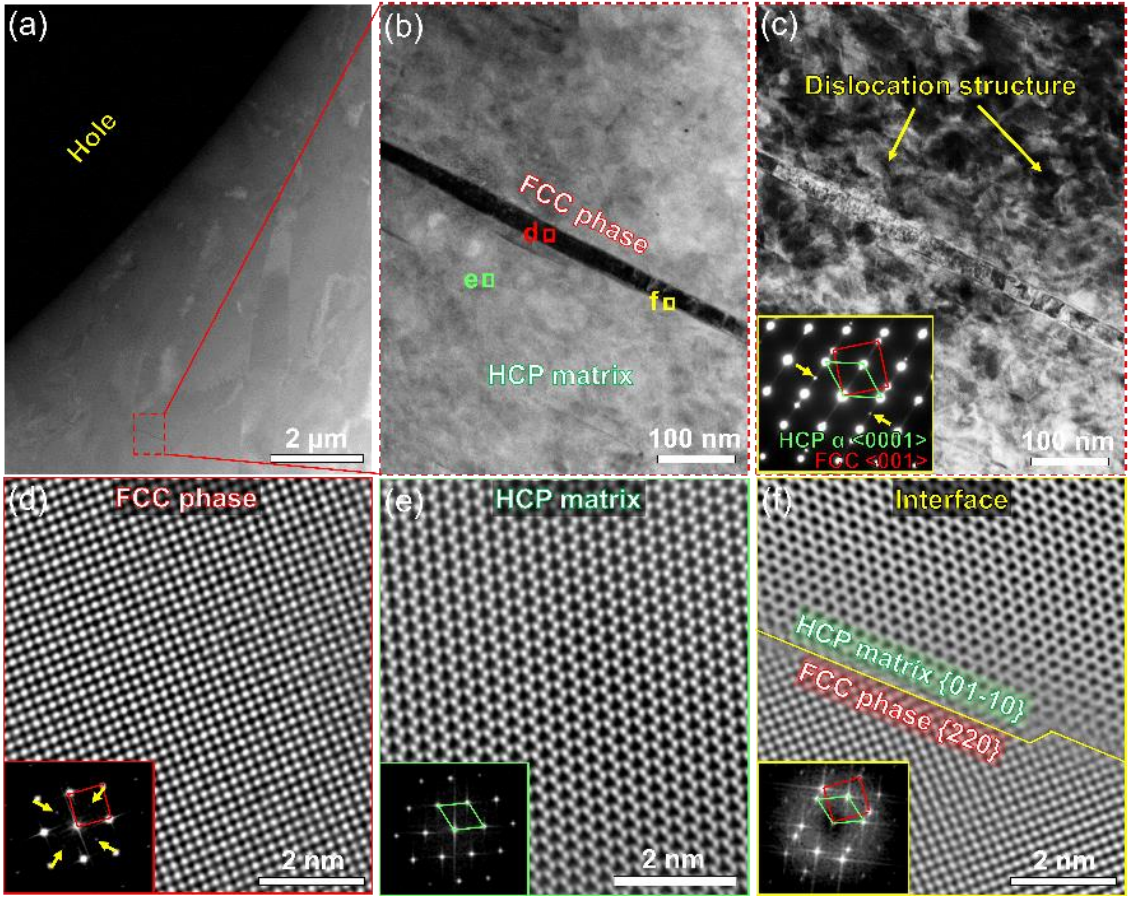

Figure 1: TEM/STEM analysis of the electropolished sample near the hole after 30% cold-rolling. The needle-shaped FCC phase is characterized at (a) low and (b) high magnification by HAADF-STEM. (c) presents the corresponding LAADF-STEM image. (d-f) High-resolution Fourier filtered STEM images corresponding to the regions delineated by the red, green, and yellow squares shown in (b), respectively. The insets are the corresponding fast-Fourier transforms. The superlattice reflections are marked by yellow arrows.

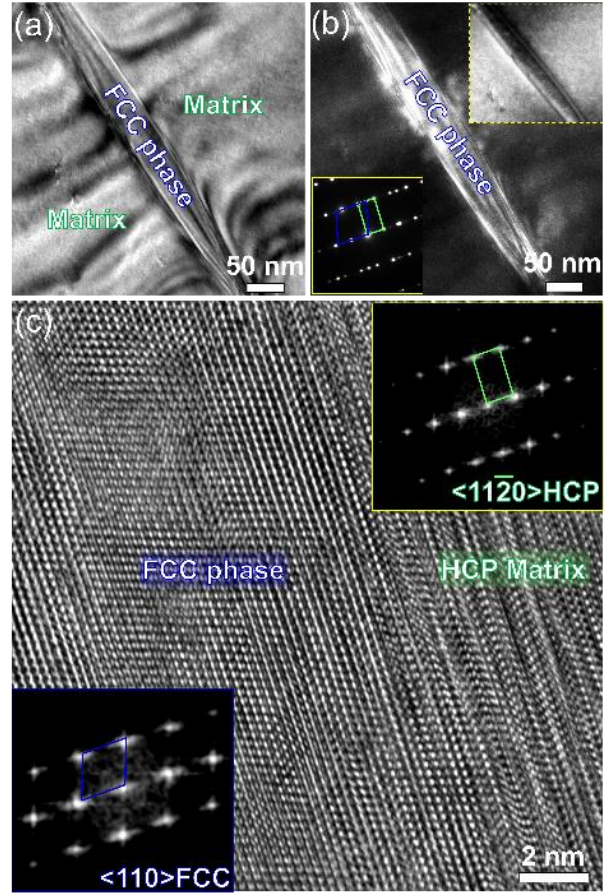

*Figure 2: The lenticular-shaped FCC phase is characterized by (a) bright-field and (b) dark-field TEM images. The inset on the bottom left is the corresponding SAED and the inset on the upper right is the HAADF-STEM image. (c) High-resolution TEM image of the interface between lenticular-shaped FCC phase and matrix. The insets are the corresponding fast-Fourier transforms for both phases.*

STEM-EELS spectrum imaging of the low-loss region was performed to determine the chemical signature of both types of FCC phases (i.e., orientation variants OR1 and OR2) found in the electropolished Ti-foil, as shown in Fig. 3a and b. The overlaid phase maps indicate that the precipitates can be clearly distinguished from the bulk phase based on the position of their plasmon peaks. The plasmon peak energy is shifted from 17.5 eV in the HCP α matrix to 19.0 eV in both FCC phases, as evidenced in Fig. 3c.

In the free electron Drude model, the plasmon energy is proportional to the square root of the free electron density, n, [57], as shown in Eq. 1,

$$E_P = \hbar e \sqrt{n/\varepsilon_0 m_e} = 1.174 \text{ eVnm}^{-\frac{3}{2}} \sqrt{n} \qquad \text{(Eq. 1)}$$

where $\hbar$, $e$, $\varepsilon_0$, $m_e$ denote the reduced Planck constant, the elementary charge, the vacuum permittivity, and the mass of an electron respectively. Therefore, an upward shift of the plasmon energy can be interpreted as an increase in the density of valence electrons, while a downward shift indicates a decrease in valence electron density. As shown in Table 1, the plasmon peak of the α-phase Ti is very close to the calculated value from the Drude model, which has also been reported in previous studies [43]. According to the Drude model, FCC-Ti with its 4 valence electrons per unit cell would have a lower plasmon energy than HCP α-Ti. The opposite trend observed from experiments indicates a higher number of valence electrons than the hypothetical FCC-Ti phase, which can be ascribed to the presence of H atoms. For example, TiH or $TiH_2$ (containing 5 or 6 valence electrons per unit cell) with a FCT/FCC lattice should have plasmon energies of 19.0 and 20.8 eV (see table 1). Our experimentally observed plasmon peak energy for the FCC phase of 19.0 eV is very close to the case of $TiH_x$ (x~1). Indeed, an upward shift of the plasmon energy with respect to the metallic matrix has been well characterized in hydrides of Ti and other hydride formers like Zr. [30,41–46]. Hence, the plasmon peak analysis clearly indicates that the needle- and lenticular-shaped FCC precipitates are hydrides, instead of pure, FCC-Ti.

Table 1. *Experimental plasmon peak energies of the HCP matrix and FCC phase, and the calculated values of Ti and their hydrides according to the Drude model*

|  | HCP | FCC/TiH/$TiH_2$ | | |
|---|---|---|---|---|
| Experimental plasmon peak energy (eV) | 17.5 | 19.0 | | |
| Volume of each Ti atom ($nm^3$) | 0.0178 | 0.0191 | | |
| Valence electrons per unit cell | 4 | 4 (FCC) | 5 (TiH) | 6 ($TiH_2$) |
| Valence electron density ($nm^{-3}$) | 224.7 | 209.9 | 262.3 | 314.8 |
| Plasmon energy (eV) | 17.6 | 17.0 | 19.0 | 20.8 |

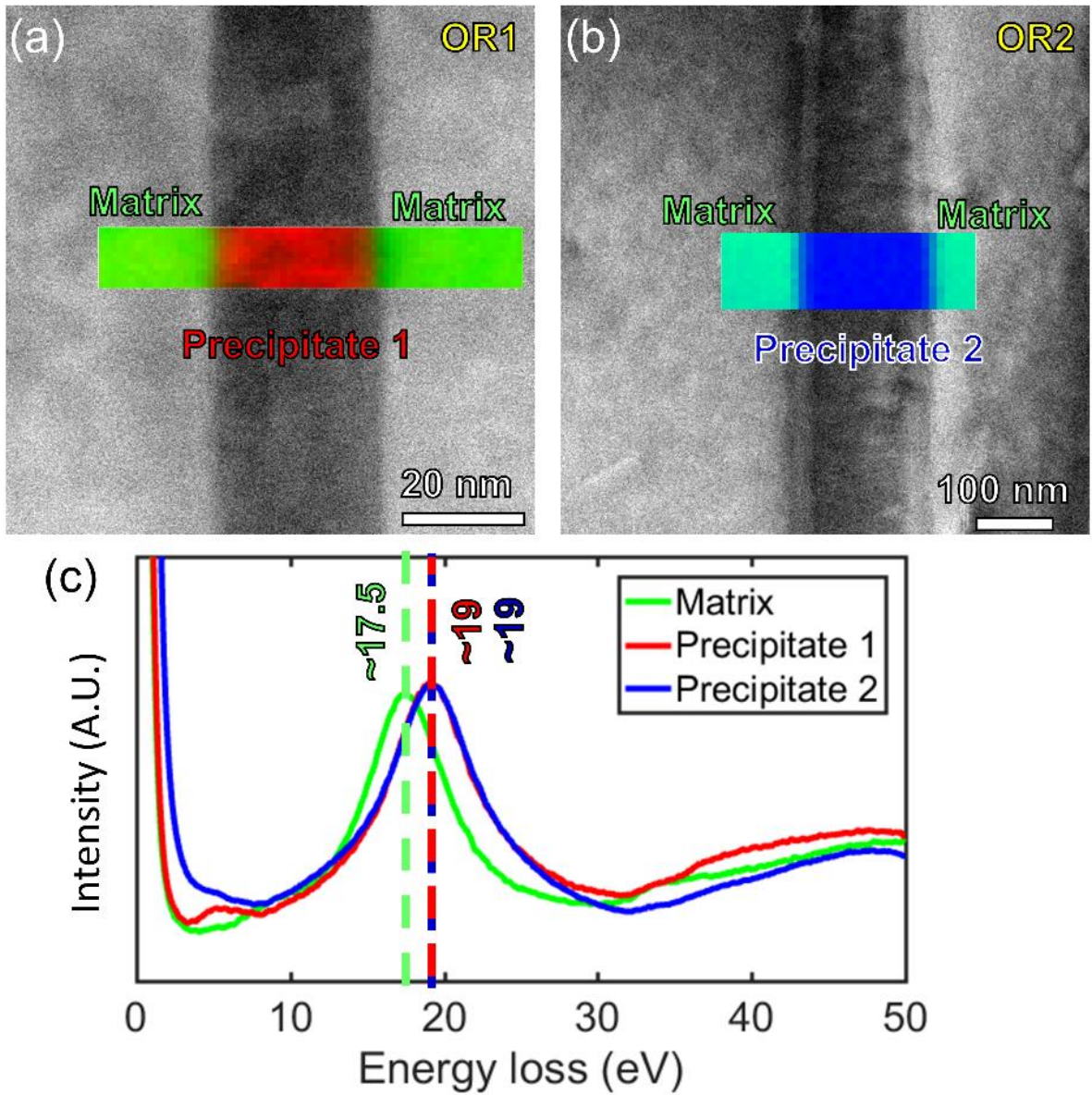

*Figure 3: (a and b) Areas of EELS spectrum imaging with overlaid phase map (green: HCP matrix, red/blue: FCC precipitates) from multivariate analysis with precipitate 1 exhibiting the OR1 orientation-relationship and precipitate 2, the OR2. (c) EELS spectra in the low energy loss region from the respective areas show their plasmon peak features at different energy loss.*

To further explore the chemical composition of the FCC phase, we conducted APT measurements on specimens extracted from thick regions of the TEM foils containing FCC-laths. For a semi-correlative APT measurement (Figs. 4a and b), a bar of the material was lifted out specifically so as to contain the FCC lath in the center. The lift-out process was carried out on a dual beam scanning electron microscope / focused ion beam (SEM/FIB) FEI Helios Plasma-FIB with a Xenon plasma source, at an accelerating voltage of 30kV and current 6–9nA at ambient temperature. The FIB cuts were made at ambient temperature, and the bar sliced and attached on the Si posts by Pt deposition at 30 keV and 48 pA ion beam current. Subsequent annular milling was conducted with 30 keV and 0.46 nA to 24 pA after the stage was cooled to -153°C [56]. The final cleaning was performed with 2 keV and 24 pA at cryogenic temperature. The APT measurements were performed on a Cameca LEAP 5000 XR, operated in high-voltage pulsing mode with 20% pulse fraction, 250 kHz pulse repetition rate and a target detection rate of 5 ions per 1000 pulses on average. The base temperature of the specimens was kept at 50 K and the pressure in the analysis chamber was consistently below $4 \times 10^{-9}$ Pa.

Figs. 4c and d show two datasets obtained from this semi-correlative APT approach. The tomographic reconstruction in Fig. 4c shows the hydrogen and Ti elemental distributions, where a hydrogen-lean and hydrogen-rich phase is clearly revealed. The 1D compositional profile along the yellow arrow shows that the hydrogen composition in the hydrogen-rich phase reaches up to approx. 50 at.%, which agrees well with previous APT studies of $TiH_x$ (x>1) by Takahashi, et al. [58] and Chang, et al. [39,59]. The ion-density map, from the segment delineated by the blue rectangle in the tomographic reconstruction, reveals the presence of two poles, identified as $\{111\}_{fcc}$ and $\{220\}_{fcc}$, in the hydrogen-rich phase, which are also typical for hydride [39,58]. Consistently, another dataset shown in Fig. 4d, with only the hydride, contains approx. 50 at.% hydrogen and exhibits the typical symmetry of the FCC crystal structure.

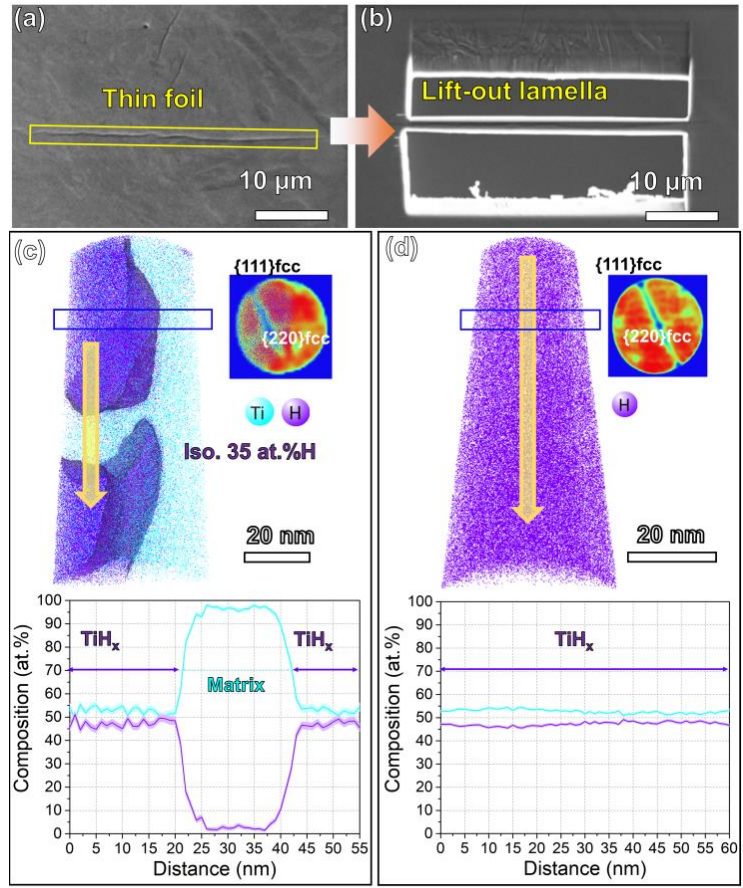

*Figure 4: Semi-correlative APT analysis of the FCC lath. (a) SEM image of FCC laths in the electropolished TEM thin foil. (b) Site-specific lift-out lamella containing the long FCC lath indicated in (a); tomographic reconstruction maps of H and Ti distribution, detector histograms exhibiting the typical symmetry of FCC crystal structure, and 1D compositional profile along the yellow arrows at (c) the interface between the hydride and the α-matrix, and (d) inside the hydride.*

The above structural and chemical analyses from both STEM-EELS plasmon peak consistently prove that the FCC phase, present in the electropolished TEM foil of cold-rolled CP-Ti in our study, is a $TiH_x$ (x~1) hydride. In addition, semi-correlative APT confirms that $TiH_x$ (x>1) hydride exists inside the same TEM foil. The lattice parameters and orientation relationship with the α matrix, as obtained by TEM analysis, agree well with earlier reports for hydrides in Ti [30,32–35]. In addition, we often found more hydride needles in the thin regions of the sample near the perforation, than in the thicker regions far from the perforation (Fig. S2), as also mentioned in other reports [19,53]. However, the exact number density of hydrides may differ in different specimens since the pick-up of H is strongly related to the specific preparation procedure for each specimen, e.g. time, area exposed to the acidic solution, sample thickness etc. Since the hydrogen content in the bulk prior to specimen preparation is only approx. 30 ppm, it is unlikely that these hydrides were initially present in the cold-rolled material. In order to further confirm that these hydrides are not native in the bulk material, but an artifact introduced during TEM specimen preparation, we prepared an electro-polished CP-Ti without cold-rolling, as shown in Fig. S3. The FCC-hydride is seen to form in the regions near the perforation, which is consistent with the observation in electro-polished CP-Ti with cold-rolling (Fig. 1). Earlier studies reported hydrogen-pick-up and hydrides and hydride formation during preparation of specimens for TEM and APT by electrochemical polishing [29,30], Ar-ion beam milling [36], or focused ion beam (FIB) at room temperature in Ti- and Zr-alloys [37–39,46], because of their high affinity for hydrogen and its low solubility in the matrix.

The intrinsic local stress concentration, associated with the cold-rolled sample, may accelerate the hydrogen accumulation and the subsequent hydride formation. The fact that more hydrides form near the perforation can first be explained by the possible higher concentration of H reached in the thinner region of the specimen due to the larger area exposed to acidic solution per unit volume. Stress relaxation normal to the surfaces of the foil also likely affects the hydride precipitation. The hydride transformation in Ti causes a volume expansion of approx. 21%, which requires both elastic and plastic accommodation [55,60], thus considerable relaxation of stress in the thin region of the foil could facilitate hydride formation. The formation of these spurious phases during specimen preparation can be misleading, in particular in the investigation of deformation mechanisms [13,18,19].

To conclude, we used STEM-EELS and APT to determine the chemistry of the FCC-type phase present in the electropolished thin foils of cold-rolled CP-Ti. The FCC phase is identified as $TiH_x$ (x≥1) hydride. These hydrides form due to the undesired yet unavoidable hydrogen pick-up during specimen preparation. Stress-relaxation during the thinning process may further promote hydrogen accumulation and accelerates hydride formation in the thin foil. Our findings highlight the importance of correlating structure and composition, and throw doubt on several reports of a new allotrope of Ti where such combined analyses were not performed.

**Acknowledgements**

Uwe Tezins, Andreas Sturm and Christian Bross are acknowledged for their support to the FIB & APT facilities at MPIE. YC is grateful to the China Scholarship Council (CSC) for funding of PhD scholarship. The authors gratefully acknowledge the fruitful discussion with Dr Ben Britton from the Department of Materials, Imperial College London. BG acknowledges financial support from the ERC-CoG-SHINE-771602.

**References**

[1]     V.L. Sliwko, P. Mohn, K. Schwartz, P. Blaha, J. Phys Condens. Matter 8 (1996) 799–815.

[2]     A. Aguayo, G. Murrieta, R. de Coss, Phys. Rev. B 65 (2002) 092106.

[3]     J. Chakraborty, K. Kumar, R. Ranjan, S.G. Chowdhury, S.R. Singh, Acta Mater. 59 (2011) 2615–2623.

[4]     Y. Sugawara, N. Shibata, S. Hara, Y. Ikuhara, J. Mater. Res. 15 (2000) 2121–2124.

[5]     F.E. Wawner, K.R. Lawless, J. Vac. Sci. Technol. 6 (1969) 588.

[6]     S.K. Kim, F. Jona, P.M. Marcus, J. Phys. Condens. Matter 8 (1996) 25–36.

[7]     D.L. Zhang, D.Y. Ying, Mater. Lett. 52 (2002) 329–333.

[8]     A.F. Jankowski, M.A. Wall, J. Mater. Res. 9 (1994) 31–38.

[9]     D. Van Heerden, D. Josell, D. Shechtman, Acta Mater. 44 (1996) 297–306.

[10]    R. Banerjee, R. Ahuja, H. l Fraser, Phys. Rev. Lett. 76 (1996) 3778–3781.

[11]    C.G. Rhodes, J.C. Williams, Metall. Trans. A 6 (1975) 1670–1671.

[12]    C.G. Rhodes, N.E. Paton, Metall. Trans. A 10 (1979) 209–216.

[13]    D.H. Hong, T.W. Lee, S.H. Lim, W.Y. Kim, S.K. Hwang, Scr. Mater. 69 (2013) 405–408.

[14]    B. Wei, S. Ni, Y. Liu, M. Song, Scr. Mater. 169 (2019) 46–51.

[15]    C.G. Rhodes, J.C. Williams, Metall. Trans. A 6 (1975) 2103–2114.

[16]    R. Jing, C.Y. Liu, M.Z. Ma, R.P. Liu, J. Alloys Compd. 552 (2013) 202–207.

[17]    Y.G. Liu, M.Q. Li, H.J. Liu, Scr. Mater. 119 (2016) 5–8.

[18]    H.C. Wu, A. Kumar, J. Wang, X.F. Bi, C.N. Tomé, Z. Zhang, S.X. Mao, Sci. Rep. 6 (2016) 1–8.

[19]    Q. Yu, J. Kacher, C. Gammer, R. Traylor, A. Samanta, Z. Yang, A.M. Minor, Scr. Mater. 140 (2017) 9–12.

[20]    J.X. Yang, H.L. Zhao, H.R. Gong, M. Song, Q.Q. Ren, Sci. Rep. 8 (2018) 1–9.

[21]    G. Han, X. Lu, Q. Xia, B. Lei, Y. Yan, C.J. Shang, J. Alloys Compd. 748 (2018) 943–952.

[22]    C. Chen, S. Qian, S. Wang, L. Niu, R. Liu, B. Liao, Z. Zhong, P. Lu, P. Li, L. Cao, Y. Wu, Mater. Charact. 136 (2018) 257–263.

[23]    X. Zheng, M. Gong, S. Xu, T. Xiong, H. Ge, Scr. Mater. 162 (2019) 1–9.

[24]    D. Josell, D. Shechtman, D. Van Heerden, Mater. Lett. 22 (1995) 275–279.

[25]    J. Bonevich, D. Josell, Phys. Rev. Lett. 82 (1999) 2002.

[26]    J. Bonevich, D. Van Heerden, D. Josell, J. Mater. Res. (1999) 1977–1981.

[27]    D. Banerjee, V.S. Arunachalam, Acta Metall. 29 (1981) 1685–1694.

[28]    D. Banerjee, Metall. Trans. A 13 (1982) 681–684.

[29]    D. Banerjee, J.C. Williams, Scr. Metall. 17 (1983) 1125–1128.

[30]    D. Banerjee, C.G. Shelton, B. Ralph, J.C. Williams, Acta Metall. 36 (1988) 125–141.

[31] D. Banerjee, Scr. Mater. 141 (2017) 146–147.

[32] H.L. Yakel, Acta Crystallogr. 11 (1958) 46–51.

[33] S.S. Sidhu, L. Heaton, D. D. Zauberis, Acta Crystallogr (1956).

[34] C. Zhang, Q. Kang, Z. Lai, Acta Metall. Mater. 42 (1994) 2555–2560.

[35] E. Conforto, D. Caillard, Acta Mater. 55 (2007) 785–798.

[36] G.J.C. Carpenter, J.A. Jackman, J.P. McCaffrey, R. Alani, Microsc. Microanal. 1 (1995) 175–184.

[37] R. Ding, I.P. Jones, J. Electron Microsc. (Tokyo). 60 (2011) 1–9.

[38] H.H. Shen, X.T. Zu, B. Chen, C.Q. Huang, K. Sun, J. Alloys Compd. 659 (2016) 23–30.

[39] Y. Chang, A.J. Breen, Z. Tarzimoghadam, P. Kürnsteiner, H. Gardner, A. Ackerman, A. Radecka, P.A.J. Bagot, W. Lu, T. Li, E.A. Jägle, M. Herbig, L.T. Stephenson, M.P. Moody, D. Rugg, D. Dye, D. Ponge, D. Raabe, B. Gault, Acta Mater. 150 (2018) 273–280.

[40] R. Ishikawa, E. Okunishi, H. Sawada, Y. Kondo, F. Hosokawa, E. Abe, Nat. Mater. 10 (2011) 278–281.

[41] B.C. Lamartine, T.W. Haas, J.S. Solomon, Appl. Surf. Sci. 4 (1980) 537–555.

[42] P. Bracconi, R. Lässer, Appl. Surf. Sci. 28 (1987) 204–214.

[43] Y. Kihn, C. Mirguet, L. Calmels, J. Electron Spectros. Relat. Phenomena 143 (2005) 117–127.

[44] N.J. Zaluzec, T. Schober, D.G. Westlake, Proc. - Annu. Meet., Electron Microsc. Soc. Am.; (United States) (1981).

[45] O.T. Woo, G.J.C. Carpenter, Scr. Metall. 20 (1986) 423–426.

[46] S.M. Hanlon, S.Y. Persaud, F. Long, A. Korinek, M.R. Daymond, J. Nucl. Mater. 515 (2019) 122–134.

[47] Y.-S. Chen, D. Haley, S.S.A. Gerstl, A.J. London, F. Sweeney, R.A. Wepf, W.M. Rainforth, P.A.J. Bagot, M.P. Moody, Science (80-. ). 355 (2017).

[48] J. Takahashi, K. Kawakami, Y. Kobayashi, Acta Mater. (2018).

[49] G. Sundell, M. Thuvander, A.K. Yatim, H. Nordin, H.-O. Andrén, Corros. Sci. 90 (2015) 1–4.

[50] I. Mouton, A.J. Breen, S. Wang, Y. Chang, A. Szczepaniak, P. Kontis, L.T. Stephenson, D. Raabe, M. Herbig, T. Ben Britton, B. Gault, Microsc. Microanal. 25 (2018) 481–488.

[51] A.J. Breen, I. Mouton, W. Lu, S. Wang, A. Szczepaniak, P. Kontis, L.T.T. Stephenson, Y. Chang, A. Kwiatkowski da Silva, C.H. Liebscher, D. Raabe, T.B. Britton, M. Herbig, B. Gault, Scr. Mater. 156 (2018) 42–46.

[52] F. Yan, I. Mouton, L.T. Stephenson, A.J. Breen, Y. Chang, D. Ponge, D. Raabe, B. Gault, Scr. Mater. 162 (2019) 321–325.

[53] Y. Chang, W. Lu, J. Guénolé, L. Stephenson, A. Szczpaniak, P. Kontis, A. Ackerman, F. Dear, I. Mouton, X. Zhong, S. Zhang, D. Dye, C.H. Liebscher, D. Ponge, S. Korte-Kerze, D. Raabe, B. Gault, Nat. Commun. (2019).

[54] S. Zhang, C. Scheu, Microscopy 67 (2018) i133–i141.

[55] H. Numakura, M. Koiwa, Acta Metall. 32 (1984) 1799–1807.

[56] L.T. Stephenson, A. Szczepaniak, I. Mouton, K.A.K. Rusitzka, A.J. Breen, U. Tezins, A. Sturm, D. Vogel, Y. Chang, P. Kontis, A. Rosenthal, J.D. Shepard, U. Maier, T.F. Kelly, D. Raabe, B. Gault, PLoS One 13 (2018) e0209211.

[57] R.F. Egerton, Electron Energy-Loss Spectroscopy in the Electron Microscope, Springer Science & Business Media, 2011.

[58] J. Takahashi, K. Kawakami, H. Otsuka, H. Fujii, Ultramicroscopy 109 (2009) 568–573.

[59] Y.H. Chang, I. Mouton, L. Stephenson, M. Ashton, G.K. Zhang, A. Szczpaniak, W.J. Lu, D. Ponge, D. Raabe, B. Gault, New J. Phys. (2019).

[60] N.E. Paton, B.S. Hickman, D.H. Leslie, Metall. Trans. 2 (1971) 2791–2796.